# Dynamic Hedging using Generated Genetic Programming Implied Volatility Models

Fathi Abid  Fathi.Abid@fsegs.rnu.tn
RU: MODESFI, Faculty of Economics and Business, Road of the Airport Km 4, 3018, Sfax, Tunisia

Wafa Abdelmalek  Wafa.Abdelmalek@fsegs.rnu.tn
RU: MODESFI, Faculty of Economics and Business, Road of the Airport Km 4, 3018, Sfax, Tunisia

Sana Ben Hamida  bhsana@yahoo.fr
RU: Laboratory of intelligent IT engineering, Higher School of Technology and Computer Science, Charguia, Tunisia

**Abstract:**
The purpose of this paper is to improve the accuracy of dynamic hedging using implied volatilities generated by genetic programming. Using real data from S&P500 index options, the genetic programming's ability to forecast Black and Scholes implied volatility is compared between static and dynamic training-subset selection methods. The performance of the best generated GP implied volatilities is tested in dynamic hedging and compared with Black-Scholes model. Based on MSE total, the dynamic training of GP yields better results than those obtained from static training with fixed samples. According to hedging errors, the GP model is more accurate almost in all hedging strategies than the BS model, particularly for in-the-money call options and at-the-money put options.

## 1. Introduction

One challenge posed by financial markets is to correctly forecast the volatility of financial securities, which is a crucial variable in trading and risk management of derivative securities. Traditional parametric methods have limited success in estimating and forecasting volatility as they are dependent on restrictive assumptions and difficult to estimate. Several machine learning techniques have been recently used to overcome these difficulties such as artificial neural networks and evolutionary computation algorithms, in particular genetic programming (GP) which is considered a promising approach to forecast financial time series (Tsang et al. (2004), Kaboudan (2005)). It has been successfully applied to forecast historical volatility (Chen and Yeh (1997), Zumbach et al. (2001), Neely and Weller (2002), Ma et al. (2006, 2007)) and implied volatility (Abdelmalek et al. (2009)).

This paper makes an initial attempt to test whether the hedger can benefit more by using generated GP implied volatilities instead of Black-Scholes implied volatilities in conducting dynamic hedging strategies.

Changes in asset prices is not the only risk faced by market participants, instantaneous changes in market implied volatility can also bring a hedging portfolio significantly out of balance. Extensive research during the last two decades has demonstrated that the volatility of stocks is not constant over time (Bollerslev et al. (1992)). Engle (1982) and Bollerslev (1986) introduced the family of ARCH and GARCH models to describe the evolution of the volatility of the asset price in discrete time. Econometric tests of these models clearly reject the hypothesis of constant volatility and find evidence of volatility clustering over time. In the financial literature, stochastic volatility models have been proposed to model these effects in a continuous-time setting (Hull and White (1987), Scott (1987), Wiggins (1987), Heston (1993)). Although these models improve the benchmark Black-Scholes model, they are complex because they require strong assumptions and computational effort to estimate parameters and stochastic process. As mentioned in Ma et al. (2004), traditional financial engineering methods based on parametric models such as the GARCH model family, seem to have difficulty to improve the accuracy in volatility forecasting due to their rigid as well as linear structure. Using its basic and flexible tree- structured representation, GP is capable of solving non-linear problems with little input or external knowledge. It has been successfully applied to forecast historical volatility (Chen and Yeh (1997), Zumbach et al. (2001), Neely and Weller (2002), Ma et al. (2006, 2007)) and implied volatility (Abdelmalek et al. (2009)).

The first thrust of this paper deals with generation of implied volatility from option markets using genetic programming. This volatility forecasting method should be free of strong assumptions regarding underlying price dynamics and more flexible than parametric methods.

Derivative asset prices are affected by new information and changes in expectations as much as they are by changes in the value of the underlying index. If traders have perfect foresight on forward volatility, then dynamic hedging would be essentially riskless. In practice, continuous hedging is impossible, but the convexity of derivative security allows for adjustments in the exposure to higher-order sensitivities of the model, such as gamma, vega… Most of the extant literature on hedging a target contract using other exchange-traded options focuses on static strategies, motivated at least in part by the desire to avoid the high costs of frequent trading. The goal of static hedging is to construct a buy-and-hold portfolio of exchange traded claims that perfectly replicates the payoff of a given over-the-counter product (Derman, Ergener and Kani (1995), Carr, Ellis and Gupta (1998)). The static hedging strategy does not require any rebalancing and therefore, it does not incur significant transaction costs. Unfortunately, the odds of coming up with a perfect static hedge for a given over-the-counter claim are small, given the limited number of exchange listed option contracts with sufficient trading volume. In other words, the static hedge can only be efficient if traded options are available with sufficiently similar maturity and moneyness as the over-the-counter product that has to be hedged.

Under a stochastic volatility, a perfect hedge can in principle be constructed with a dynamically rebalanced portfolio consisting of the underlying and one additional option. In practice, the dynamic replication strategy for European options will only be perfect if all of the assumptions underlying the Black-Scholes formula hold. For general contingent claims on a stock, under market frictions, the delta might still be used as first-order approximation to set up a riskless portfolio. However, if the volatility of the underlying stock varies stochastically, then the delta hedging method might fail severely. A simple method to limit the volatility risk is to consider the volatility sensitivity vega of the contract. The portfolio will have to be rebalanced frequently to ensure delta-vega neutrality. With transaction costs, frequent rebalancing might result in considerable losses. In practice, investors can rebalance their portfolios only at discrete intervals of time to reduce transactions costs.

Non parametric hedging strategies as an alternative to the existing parametric model based-strategies, have been proposed by Hutchinson et al (1994) and Aït-Sahalia and Lo (1998). Those studies estimated pricing formulas by nonparametric or semi-parametric statistical methods such as neural networks and kernel regression, and they measured their performance in terms of delta-hedging. Few researches have focused on the dynamic hedging using genetic programming, however. Chen et al. (1999) have applied genetic programming to price and hedge S&P500 index options. By distinguishing the case in-the-money from the case out-of-the-money, the performance of GP is compared with the Black-Scholes model in terms of its hedging accuracy. Based on the post-sample performance, it is found that in approximately 20% of the 97 test paths, GP has lower tracking error than the BS formula.

The second thrust of this paper is to study the accuracy of the generated GP implied volatility models in terms of dynamic hedging.
Since the true volatility is unobservable, it is impossible to assess the accuracy of any particular model; forecasts can only be related to realized volatility. The strategy adopted in this paper is to assume that the implied volatility is a reasonable proxy for realized volatility, to generate forecasting implied volatility models using genetic programming and then to analyze the implications of this predictability for hedging purposes.

The rest of the paper is organized as follows: section 2 provides background information regarding genetic programming approach, section 3 describes research design and methodology used in this paper, section 4 reports experimental results and finally section 5 concludes.

## 2. Genetic programming background

Genetic Programming (Koza (1992)) is an evolutionary algorithm that attempts to evolve a population of solutions to a problem according to the Darwinian principles of natural evolution. It is an extension of the basic genetic algorithms as introduced by Holland (1975). The major difference between genetic programming and genetic algorithms is the representation of the solution candidates. A hierarchical tree structure represents the solution candidates in genetic programming while a string of characters with a fixed length, called chromosome, represents the solution candidates in genetic algorithms. This makes the GP a flexible technique as the solutions can vary in size and shape. An interesting feature of GP is its ability to perform optimization at a structural level. This is an attractive prospect as the algorithm can simultaneously evolve a model's function form and numerical parameter values. The genetic programming's algorithm structure consists of the following elements: nodes definition, initialization, fitness evaluation, selection, genetic operators (crossover and mutation) and termination condition.

*Nodes Definition*: The nodes in the tree structure of genetic programming can be classified into terminal (leaf) nodes and function (non-terminal) nodes. The terminal set, which corresponds to the inputs of the program, is determined according to the domain of problems and the elements can be constants or variables. The function set may be standard arithmetic operations, standard programming operations, standard mathematical functions, logical functions, or domain-specific functions.
The terminal set and the function set are used to construct trees which are solutions to the problem.
*Initialization*: Genetic programming starts with an initial population which is a set of randomly generated trees.
*Fitness function*: The evolutionary process is driven by a fitness that evaluates how well each individual performs in the problem environment.
*Selection*: The selection method determines how to select individuals from the population to be parents for next generation based on fitness function. Parents with better quality are usually selected with the hope that they can produce better offspring with larger chance.
*Crossover*: The crossover operation creates new offspring trees from two selected parents by exchanging their sub-trees.
*Mutation*: The mutation operator randomly changes a tree by randomly altering nodes or sub-trees to create a new offspring.
*Termination Condition*: The termination conditions for genetic programming usually include the convergence level of evolution process or the maximum number of generations.

## 3. Research design and methodology

Dynamic hedging is very sensitive to volatility forecast and good hedges require accurate estimate of volatility. Implied volatilities, generated from option markets, can be particularly useful in such contents as they are forward-looking measures of the market's expected volatility during the remaining life of an option (Blair et al. (2001), Bush et al. (2007)…). This paper proposes a non parametric approach based on GP algorithm to improve the accuracy of the implied volatility forecast and consequently the dynamic hedging.

In the standard GP, the entire population of GP function-trees is evaluated against the entire training data set, so the number of function-tree evaluations carried out per generation is directly proportional to both the population size and the size of the training set. GP can encounter the problem of managing training sets which are too large to fit into the memory of computers, and then the realization of predictors. In this case, data reduction through the partitioning of the data set into smaller subsets seems a good approach. In this paper, sample data are split into times series and moneyness-time to maturity classes.

GP learns from the training set. Test set is used to evaluate its performance during and after training. As data are divided in several sub samples, two training-subset selection methods are used. Abdelmalek et al. (2009) proposed a static training approach allowing the GP to learn separately on different training sub samples. Such approach might provide local solutions not adaptive to the entire enlarged data set. Alternatively, a dynamic training approach is developed. It allows GP to learn simultaneously on all training sub samples and it implies a new parameter added to the basic GP algorithm which is the number of generations to change sample. This approach lightens the training task for the GP and favors the discovery of solutions that are more robust across different learning data samples and seem to have better generalization ability. Comparative experiments are provided to show how successfully dynamic training subset-selection methods are applied to improve the robustness of GP to generate general models relative to static training-subset selection method. The best forecasting implied volatility models are selected according to total mean squared error (MSE) criterion.

Accurate volatility forecasting is an essential element in conducting good dynamic hedging strategies. The best generated GP implied volatilities are used to compute hedge factors and implement delta-neutral, delta-gamma neutral and delta-vega neutral strategies. According to the average hedging error, the GP hedging performance is compared to that of Black-Scholes model and the main results are reported.

Figure 1 illustrates the operational procedure to implement the proposed approach.

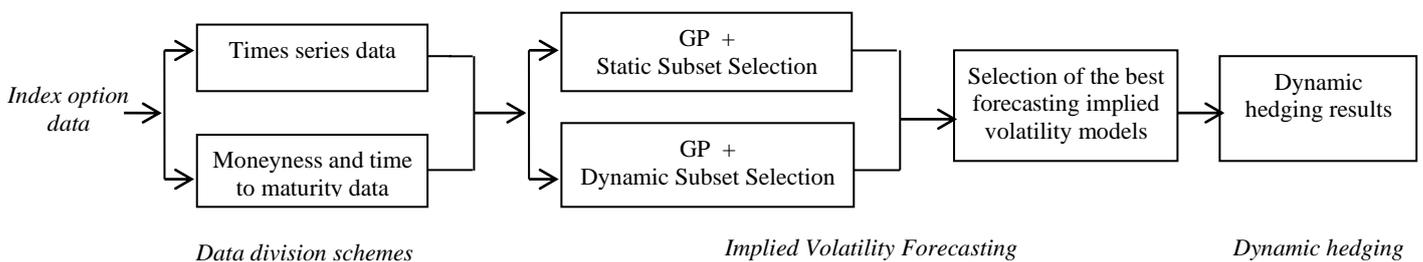

*Figure 1: Description of the proposed approach's implementation*

The operational procedure consists of the following steps: The first step is devoted for the data division schemes. The second step deals with the implementation of GP[1], the application of training subset selection methods and the selection of the best forecasting implied volatility models. The last step is dedicated to dynamic hedging results.

---
[1] GP system is built around the Evolving Object library, which is an *ANSI-C++ evolutionary computation Framework* (EO library).

## 3.1. Data division schemes

Data used in this study consist of daily prices for the European-style S&P 500 index calls and puts options traded on the Chicago Board of Options Exchange from 02 January to 29 August 2003. The data base include the time of the quote, the expiration date, the exercise price and the daily bid and ask quotes for call and put options. Similar information for the underlying S&P 500 index is also available on a daily basis. S&P500 index options are among the most actively traded financial derivatives in the world. The minimum tick for series trading below 3 is 1/16 and for all other series 1/8. Strike price intervals are 5 points, and 25 for far months. The expiration months are three near term months followed by three additional months from the March quarterly cycle (March, June, September, and December). Following a standard practice, we use the average of an option's bid and ask price as a stand-in for the market value of the option. The risk free interest rate is approximated by using 3 month US Treasury bill rates. It is assumed that there are no transaction costs and no dividend.

To reduce the likelihood of errors, data screening procedures are used (Harvey and Whaley (1991, 1992)). We apply four exclusion filters to construct the final option sample. First, as implied volatilities of short-term options are very sensitive to small errors in the option price and may convey liquidity-related biases, options with time to maturity less than 10 days are excluded. Second, options with low quotes are eliminated to mitigate the impact of price discreteness on option valuation. Third, deep-ITM and deep-OTM option prices are also excluded due to the lack of trading volume. Finally, option prices not satisfying the arbitrage restriction (Merton (1973)), $C \geq S - Ke^{-r\tau}$, are not included.

The final sample contains 6670 daily option quotes, with ATM, ITM and OTM options respectively taking up 37%, 34% and 29% of the total sample.

In this paper, two data division schemes are used. The full sample is sorted first, by time series (TS) and second by moneyness-time to maturity (MTM). For time series, data are divided into 10 successive samples ($S_1$, $S_2$…$S_{10}$), each contains 667 daily observations. The first nine samples are used as training sub samples. For moneyness-time to maturity, data are divided into nine classes with respect to moneyness and time to maturity criteria. According to moneyness criterion: A call option is said OTM if $S/K < 0.98$; ATM if $S/K \in [0.98, 1.03[$; and ITM if $S/K \geq 1.03$. According to time to maturity criterion: A call option is Short Term (ST) if $\tau < 60$ days; Medium Term (MT) if $\tau \in [60, 180]$ days; and Long Term (LT) if $\tau > 180$ days. Each class $C_i$ is divided on training set $C_i^L$ and test set $C_i^T$, which produces respectively nine training and nine test moneyness- time to maturity sub classes. Figure 2 illustrates the two division schemes.

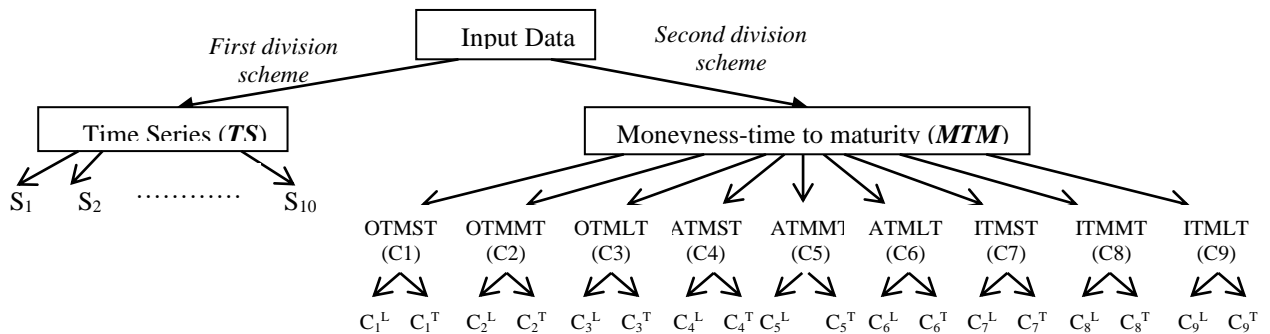

*Figure 2: Data division schemes*

## 3.2. Implied volatility forecasting using GP:

The implied volatility is defined as the standard deviation which equates the model option price to the observed market option price. Since there is no explicit formula available to compute directly the implied volatility, the latter can be obtained by inverting the option pricing model. On the contrary, the GP offers explicit formulas which can compute directly the implied volatility. Therefore, the GP is applied to forecast implied volatility from the S&P500 index options.

This subsection will describe the experiments that have been accomplished using the GP algorithm to forecast implied volatility. In the first experiment, the GP is trained using static training-subset selection method; in the second one, we used dynamic training-subset selection methods. We will describe training and test samples that were used in the experiments and the GP parameters setting.

### 3.2.1. Training subset selection methods:

In this paper, static and dynamic training-subset selection approaches are used. First, the GP is trained independently on each sub sample relative to each data division scheme (Abdelmalek et al. (2009)). This approach is called static training-subset selection method. The following flow chart summarizes the GP's algorithm structure using static training-subset selection method.

*Initialize population*
***while*** *(termination condition not satisfied)* ***do***
***begin***
    *Evaluate the performance of each individual according to the fitness criterion*
    *Select individuals in the population using the selection algorithm*
    *Perform crossover and mutation on the selected individuals*
    *Replace the existing population by the new population*
***endwhile***
*Report the best solution found*
***end***

***Algorithm 1: A flow chart summarizing the GP's algorithm structure using static training-subset selection method.***

Second, the GP is trained simultaneously on the entire data sub samples relative to each data division scheme, rather than just a single subset by changing the training sub sample during the run process. This approach is called dynamic training- subset selection method. The main goal of this method is to make GP adaptive to all training samples and able to generate general models and solutions that are more robust across different learning data samples. Four dynamic training-subset selection methods are applied: *Random Subset Selection* method (*RSS*), *Sequential Subset Selection* method (*SSS*), *Adaptive-Sequential Subset Selection* method (ASSS) and *Adaptive-Random Subset Selection* method (*ARSS*). The *RSS* and *SSS* allow the GP to learn on all training samples in turn (*SSS*) or randomly (*RSS*). However, with these methods, there is no certainty that GP will focus on the samples which are difficult to learn. Then, the *ASSS* and the *ARSS,* which are variants of *the adaptive subset selection (ASS),* are introduced to focus the GP's attention onto the difficult samples i.e. having the greatest MSE and then to improve the learning algorithm.

The following flow chart summarizes the GP's algorithm structure using dynamic training-subset selection method.

```
Initialize population
t=1; (t: current generation)
While (termination condition not satisfied) do
Begin
      Select the first training sample s_i according to the selection method (RSS,SSS, ARSS,ASSS)
 Until the generations number to change sample is reached
         - Select individuals in the population using the selection algorithm
         - Perform crossover and mutation operations on the selected individuals
         - Evaluate the performance of individuals according to the fitness criterion
  Select next training sample s_i
  End while
  Report the best solution
End
```

***Algorithm 2: A flow chart summarizing the GP's algorithm structure using dynamic training-subset selection method.***

Let $S= \{S_1, S_2,...,S_k\}$ be the set of training samples $S_{i\ (i=1...k)}$, where k is the total number of samples. A selection probability P ($S_i$) is allocated to each sample $S_i$ from S. The training sample $S_i$ is changed each g generation (g is the number of generations to change sample) according to this selection probability and the dynamic training-subset selection method used. This procedure is repeated until the maximum number of generations is reached. This permits GP to adapt its generating process to changing data in response to feedback from the fitness function.

   a- *Random training-Subset Selection method (RSS)*:
It selects randomly the training samples with replacement. At each g generation, all the samples from S have the same probability to be selected as the current training sample:
P ($S_i$) =1/k, 1≤ i ≤ k.

   b- *Sequential training-Subset Selection method (SSS)*
It selects all the training samples in the order. All the learning subsets are used during the evolution in an iterative way.
If, at generation g-1, the current training sample is $S_i$, then at generation g:

$$P(S_j) = 1, \text{ avec } j= i+1 \text{ si } i<k, \text{ ou } j=1 \text{ si } i=k$$

   c- *Adaptive training-Subset Selection method (ASS):*
Instead of selecting a training subset data in a random or sequential way, one can use an adaptive approach to dynamically select difficult training subsets data which are frequently misclassified. This approach is inspired from the dynamic subset selection method proposed by Gathercole and Ross (1994) which is based on the idea of dynamically selecting instances, not training samples, which are difficult and/or have not been selected for several generations. Selection is made according to a weight computed proportionally to the sample's average fitness. Each g generations, the weights are updated as follows:

$$W(S_i) = \frac{\sum_{t=1}^{g} \sum_{j=1}^{M} f(X_j)}{M * g} \quad (3)$$

Where, M is the size of $S_i$ ( $X_j \in S_i$ ), g is the number of generations to change sample, and $f(X_j)$ is the MSE of the individual $X_j$.

Selection of the training sample for the next run is made according to its misclassification measured by its higher fitness error. At each g generations, training samples are re-ordered, so that the most difficult training samples, which have higher fitness errors, will be moved to the beginning of the ordered training list, and the easiest training samples, which have smaller fitness errors, will be moved to the end of the ordered training list.

*c-1. Adaptive-Sequential training-Subset Selection method (ASSS):*

The initial weights are initialized with a constant and the selection of samples is made in an iterative way: $W(S_i) = C$, $1 \leq i \leq k$.

*c-2. Adaptive-Random training-Subset Selection method (ARSS):*

The initial weights are generated randomly in the start of running, rather than initialized with a constant.

$$\text{For } t=1, \ W(S_i) = \tilde{P}_i, \tilde{P}_i \in [0,1], 1 \leq i \leq k$$

### 3.2.2. Training and test samples

Different forecasting GP volatility models are estimated from the training set and judged upon their performance on the test set. Table 1 summarizes the training and test data samples used for static and dynamic training-subset selection methods, respectively.

| Subset Selection | Learning data sample | Test data sample |
|---|---|---|
| Static Subset Selection | 1. $S_i \in$ TS samples ($S_1, \ldots, S_9$) *(1 subset for a run)* | The successive TS sample $S_j$, j=i+1 |
| | 2. $C_i^L \in$ MTM training samples ($C_1^L, \ldots, C_9^L$)  *(1 subset for a run)* | The corresponding MTM test samples $C_i^T$ |
| Dynamic Subset Selection (RSS/SSS/ASSS/ARSS) | 1. TS samples $S_1, \ldots, S_9$ *(9 subsets for a run)* | The last subset in TS samples set ($S_{10}$) |
| | 2. MTM training samples $C_1^L, \ldots, C_9^L$  *(9 subsets for a run)* | The nine MTM test samples ($C_1^T + C_2^T \ldots + C_9^T$) |
| | 3. TS samples + MTM samples ($S_1, \ldots, S_9, C_1^L, \ldots, C_9^L$ ) *(18 subsets for a run)* | The last TS sample with the nine MTM test samples ($S_{10} + C_1^T + C_2^T \ldots + C_9^T$) |

*Table 1: Definition of training and test data samples for static and dynamic training-subset selection methods.*

In static training-subset selection approach, first, the genetic program is trained separately on each of the first nine time series sub samples ($S_1, \ldots, S_9$) using ten different seeds and is tested on the subset data from the immediately following date ($S_2, \ldots, S_{10}$). Second, using the same genetic parameters and random seeds applied for time series data, the GP is trained separately on each of the first nine moneyness- time to maturity sub classes ($C_1^L, \ldots, C_9^L$) and is tested on the second nine moneyness- time to maturity sub classes ($C_1^T, \ldots, C_9^T$) .

In dynamic training-subset selection approach, first, the genetic program is trained on the first nine time series sub samples simultaneously ($S_1, \ldots, S_9$) using ten different seeds and it is tested only on the tenth sub sample data ($S_{10}$). Second, the GP is trained on the first nine moneyness- time to maturity sub classes simultaneously  ($C_1^L, \ldots, C_9^L$) and it is tested on the second nine moneyness- time to maturity sub classes regrouped in one test sample data ($C_1^T + C_2^T \ldots + C_9^T$). Third, the GP is trained on both the nine time series sub samples and the nine moneyness- time to maturity sub classes simultaneously ($S_1, \ldots, S_9, C_1^L, \ldots, C_9^L$ ) and it is tested on one test sample data composed of the time series and moneyness- time to maturity test data ($S_{10} + C_1^T + C_2^T \ldots + C_9^T$).

### 3.2.3. Parameters setting

Our GP software is referred to as symbolic regression written in C++ language. It is designed to find a function that relates a set of inputs to an output without making any assumptions about the structure of that function. Symbolic regression was one of the earliest applications of GP (Koza, 1992), and has continued to be widely studied (Cai et al. (2006); Gustafson et al. (2005); Keijzer (2004); Lew et al. (2006)).

The terminal and function sets used are described in Table 2. The terminal set includes the inputs variables, notably, the option price divided by strike price ($\frac{C}{K}$ for calls and $\frac{P}{K}$ for puts), the index price divided by strike price $\frac{S}{K}$ and time to maturity $\tau$. The function set includes unary and binary nodes. Unary nodes consist of mathematical functions, notably, cosinus function (cos), sinus function (sin), log function (ln), exponential function (exp), square root function ($\sqrt{}$) and the normal cumulative distribution function ($\Phi$). Binary nodes consist of the four basic mathematical operators, notably, addition (+), subtraction (-), multiplication ($\times$) and division ($\%$). The basic division operation is protected against division by zero and the log and square root functions are protected against negative arguments.

|  | Expression | Definition |
|---|---|---|
| Terminal Set | C/K | Call price / Strike price |
|  | S/K | Index price / Strike price |
|  | $\tau$ | Time to maturity |
| Function Set | + | Addition |
|  | - | Subtraction |
|  | * | Multiplication |
|  | % | Protected division: x % y = 1 if y=0; x % y = x % y otherwise |
|  | ln | Protected natural log: $\ln(x) = \ln(|x|)$ |
|  | Exp | Exponential function: $\exp(x) = e^x$ |
|  | Sqrt | Protected square root: $\sqrt{x} = \sqrt{|x|}$ |
|  | Ncdf | Normal cumulative distribution function $\Phi$ |

*Table 2: Terminal set and function set.*

The Black-Scholes implied volatility $\sigma_t^{BS}$ is used as target output. It is defined as the standard deviation which equates the BS model price $C_{BS}$ [2] to the market option price $C_t^*$:

$$\exists! \sigma_t^{BS}(K,T) \succ 0,$$
$$C_{BS}\left(S_t, K, \tau, \sigma_t^{BS}(K,T)\right) = C_t^*(K,T) \quad (1)$$

The generated GP trees provide at each time t the forecast value $\hat{\sigma}_t$, and the fitness function used to measure the accuracy of forecast is the mean squared error (MSE) computed as follows:

$$MSE = \frac{1}{N} \sum_{t=1}^{N} \left(\sigma_t^{BS} - \hat{\sigma}_t\right)^2 \quad (2)$$

Where, N is the number of data sample.

---

[2] ($C_{BS} = SN(d_1) - Ke^{-r\tau} N(d_2)$, $d_1 = \dfrac{\ln\left(\dfrac{S}{K}\right) + \left(r + 0.5\sigma^2\right)\tau}{\sigma\sqrt{\tau}}$, $d_2 = d_1 - \sigma\sqrt{\tau}$).

The implementation of genetic programming involves a series of trial and error experiments to determine the optimal set of genetic parameters which is listed in Table 3.

| | |
|---|---|
| Population size: | 100 |
| Offspring size: | 200 |
| Maximum number of generations for static method: | 400 |
| Maximum number of generations for dynamic method: | 1000 |
| Generations' number to change sample | 20-100 |
| Maximum depth of new individual: | 6 |
| Maximum depth of the tree: | 17 |
| Tournament size: | 4 |
| Crossover probability: | 60% |
| Mutation probability: | 40% |
|     Branch mutation: | 20% |
|     Point mutation: | 10% |
|     Expansion mutation: | 10% |

*Table 3: Summary of GP parameters.*

The generated GP volatility models are performed using a ramped half and half as initialization method (Koza (1992)). This method involves generating an equal number of trees using a maximum initial depth that ranges from 2 to 6. For each level of depth, 50% of the initial trees are generated via the full method and the other 50% are generated via the grow method. A maximum size of tree measured by depth is 17. This is a popular number used to limit the size of tree (Koza (1992)). It is large enough to accommodate complicated formulas and works in practice. Based on the fitness criterion, the selection of the individuals for reproduction is done with the tournament selection algorithm. A group of individuals is selected from the population with a uniform random probability distribution. The fitness values of each member of this group are compared and the actual best is selected. The size of the group is given by the tournament size which is equal here to 4. The crossover operator is used to generate about 60% of the individuals in the population, while the mutation operator is used to generate about 40% of the population. Different mutation operators are used. Point mutation operator consists of replacing a single node in a tree with another randomly-generated node of the same arity. Branch mutation operator randomly selects an internal node in the tree, and then it replaces the subtree rooted at that node with a new randomly-generated subtree. Expansion mutation operator randomly selects a terminal node in the tree, and then replaces it with a new randomly-generated subtree. Branch mutation is applied with a rate of 20%; Point and Expansion mutations are applied with a rate of 10% each. The method of replacing parents for the next generation is comma replacement strategy, which selects the best offspring to replace the parents. It assumes that offspring size is higher than parents' size. The stopping criterion is the maximum number of generations. It is fixed at 400 and 1000 for static and dynamic training- subset selection, respectively. In the dynamic training- subset selection approach, the maximum number of generations is increased to allow the GP to train on the maximum of samples simultaneously. The number of generations to change sample varied between 20 and 100 generations.

Based on the training and test MSE, the best generated GP volatility models relative to static and dynamic training-subset selection methods respectively are selected. These models are then compared with each other according to the MSE total and the best ones are used to implement the dynamic hedging strategies as described in the following section.

### 3.3. Dynamic hedging:

To assess the accuracy of selected generated GP volatility models in hedging with respect to Black-Scholes model, three dynamic hedging strategies are employed, notably, delta-neutral, delta-gamma neutral and delta-vega neutral strategies.

For delta hedging, at date zero, a delta hedge portfolio consisting of a short position in one call (or put) option and a long (short) position in the underlying index is formed. At any time t, the value of the delta hedge portfolio $P(t)$ is given by:

$$P(t) = V(t) + \Delta_V(t)S(t) + \beta(t) \quad (4)$$

Where, $P(t), V(t), S(t), \Delta_V(t)$ and $\beta(t)$ denote the values of the portfolio, hedging option (call or put), underlying, delta hedge factor and bond (money market account) respectively.

The portfolio is assumed self-financed, so the initial value of the hedge portfolio at the beginning of the hedge horizon is zero:

$$P(0) = V(0) + \Delta_V(0)S(0) + \beta(0) = 0 \quad (5)$$

$$\Rightarrow \beta(0) = -(V(0) + S(0)\Delta_V(0)) \quad (6)$$

A dynamic trading strategy is performed in underlying and bond to hedge the option during the hedge horizon. The portfolio rebalancing takes place at intervals of length $\delta t$ during the hedge horizon $[0, \tau], o < \tau \leq T$, where $T$ is the maturity of the option. At each rebalancing time $t_i$, the hedge factor $\Delta_V(t_i)$ is recomputed and the money market account is adjusted:

$$\beta(t_i) = e^{r\delta t}\beta(t_{i-1}) - S(t_i)(\Delta_V(t_i) - \Delta_V(t_{i-1})) \quad (7)$$

The delta hedge error is defined as the absolute value of the delta hedge portfolio at the end of the hedge horizon of the option, $|P(\tau)|$.

For delta-gamma hedging, a new position in a traded option is required. Then, the delta-gamma hedge portfolio is formed with:

$$P(t) = V(t) + x(t)S(t) + y(t)V_1(t) + B(t) \quad (8)$$

Where, $V_1(t)$ is the value of an additional option which depends on the same underlying, with the same maturity but different strike price than the hedging option $V(t)$. $x(t)$ and $y(t)$ are the proportions of the underlying and the additional option respectively. They are chosen such that the portfolio $P(t)$ is both delta and gamma neutral:

$$\begin{cases} Delta \text{ neutral}: \Delta_V(t) + x(t) + y(t)\Delta_{V1}(t) = 0 \\ Gamma \text{ neutral}: \Gamma_V(t) + y(t)\Gamma_{V_1}(t) = 0 \end{cases} \quad (9)$$

$$\Rightarrow \begin{cases} y(t) = \dfrac{-\Gamma_V(t)}{\Gamma_{V_1}(t)} \\ x(t) = -\Delta_V(t) - y(t)\Delta_{V1}(t) \end{cases} \quad (10)$$

Where, the values of $\Delta_V(t)$ and $\Gamma_V(t)$ are the delta and gamma factors for the option $V(t)$; the values $\Delta_{V_1}(t)$ and $\Gamma_{V_1}(t)$ are the delta and gamma factors for the option $V_1(t)$.

At the beginning of the hedge horizon, the value of the hedge portfolio is zero:

$$P(0) = V(0) + x(0)S(0) + y(0)V_1(0) + B(0) = 0 \quad (11)$$

$$\Rightarrow B(0) = -(V(0) + x(0)S(0) + y(0)V_1(0)) \quad (12)$$

At each rebalancing time $t_i$, both delta and gamma hedge factors are recomputed and the money market account is adjusted:

$$B(t_i) = e^{r\delta t} B(t_{i-1}) - (x(t_i) - x(t_{i-1}))S(t_i) - (y(t_i) - y(t_{i-1}))V_1(t_i) \qquad (13)$$

The delta-gamma hedge error is defined as the as the absolute value of the delta-gamma hedge portfolio at the end of the hedge horizon of the option, $|P(\tau)|$.

For delta-vega hedging, a new position in a traded option is required as in the delta-gamma hedging. The proportions of the underlying $x(t)$ and the additional option $y(t)$ are chosen such that the portfolio $P(t)$ is both delta and vega neutral:

$$\begin{cases} Delta\ neutral: \Delta_V(t) + x(t) + y(t)\Delta_{V_1}(t) = 0 \\ Vega\ neutral: \vartheta_V(t) + y(t)\vartheta_{V_1}(t) = 0 \end{cases} \qquad (14)$$

$$\Rightarrow \begin{cases} y(t) = \dfrac{-\vartheta_V(t)}{\vartheta_{V_1}(t)} \\ x(t) = -\Delta_V(t) - y(t)\Delta_{V_1}(t) \end{cases} \qquad (15)$$

Where, $\vartheta_V(t)$ and $\vartheta_{V_1}(t)$ are the vega factors for the options $V(t)$ and $V_1(t)$ respectively.

As in delta-gamma hedging, at each rebalancing time $t_i$, both delta and vega hedge factors are recomputed and the money market account is adjusted. The delta-vega hedge error is defined as the as the absolute value of the delta-vega hedge portfolio at the end of the hedge horizon of the option, $|P(\tau)|$.

35 option contracts are used as hedging options and similarly 35 contracts which depend on the same underlying, with the same maturity but different strike prices than the hedging options are used as additional options. Contracts used to implement the hedging strategies are divided according to moneyness and time to maturity criteria, which produces nine classes.

The delta, gamma and vega hedge factors are computed using the BS formula by taking the derivative of option value with respect to index price, the derivative of delta with respect to index price and the derivative of option value with respect to volatility respectively. For the GP models, the hedge ratios are computed using the same formulas replacing the BS implied volatilities with the generated GP volatilities.

Two rebalancing frequencies are considered: 1-day and 7 days revision.

The average hedging error is used as performance measure. For a particular moneyness- time to maturity class, the tracking error is given by:

$$\begin{cases} \varepsilon_M = \dfrac{\sum_{i=1}^{n}\varepsilon_i(\tau)}{n} \\ \varepsilon_i = e^{-rT} \times \dfrac{|P_i(\tau)|}{N \times V(0)} \end{cases} \qquad (16)$$

Where, n is the number of options corresponding to a particular moneyness-time to maturity class and $\varepsilon_i(\tau)$ is the present value of the absolute hedge error of the portfolio $|P(\tau)|$ over the observation path N (as a function of rebalancing frequency), divided by the initial option price $V(0)$.

## 4. Result Analysis and empirical findings:
### 4.1. Selection of the Best Genetic Programming-Implied Volatility Forecasting Models

Selection of the best generated GP volatility model, relative to each training set, for time series, moneyness-time to maturity, and both time series and moneyness-time to maturity classifications, is made according to the training and test MSE. For static training- subset selection method, nine generated GP volatility models are selected for time series (M1S1…M9S9) and similarly nine generated GP volatility models are selected for moneyness-time to maturity classification (M1C1…M9C9). The performance of these models is compared according to the MSE Total, computed using the same formula as the basic MSE for the enlarged data sample.

Table 4 reports the MSE total and the standard deviation (in parentheses) of the generated GP volatility models, using static training-subset selection method, relative to the time series samples (*TS*) and the moneyness-time to maturity classes (*MTM*).

| TS Models | MSE Total | MTM Models | MSE Total |
|---|---|---|---|
| M1S1 | 0,002723 (0,004278) | M1C1 | 256,656757 (20606,1723) |
| M2S2 | 0,005068 (0,006213) | M2C2 | 0,006921 (0,03220954) |
| M3S3 | 0,003382 (0,004993) | M3C3 | 0,030349 (0,07619638) |
| **M4S4** | **0,001444 (0,002727)** | **M4C4** | **0,001710 (0,0046241)** |
| M5S5 | 0,002012 (0,003502) | M5C5 | 1,427142 (33,3651158) |
| M6S6 | 0,001996 (0,003443) | **M6C6** | **0,002357 (0,0040963)** |
| M7S7 | 0,001901 (0,003317) | M7C7 | 0,261867 (0,30325633) |
| M8S8 | 0,002454 (0,004005) | M8C8 | 0,004318 (0,0084793) |
| M9S9 | 0,002419 (0,004095) | M9C9 | 0,002940 (0,01049067) |

*Table 4: Performance of the generated GP volatility models using static training-subset selection method, according to MSE total for the time series samples (TS) and the moneyness-time to maturity classes (MTM)*

Table 4 shows that, for the time series samples, the generated GP model M4S4 has the smallest MSE in enlarged sample. For the moneyness-time to maturity classes, the generated GP volatility models M4C4 and M6C6 seem to be more accurate in forecasting implied volatility than the other models as they present near MSE on the enlarged sample.

According to the MSE total, the time series model M4S4 seems to be more performing than moneyness-time to maturity models M4C4 and M6C6 for the enlarged sample.

For dynamic training-subset selection methods (RSS, SSS, ASSS and ARSS), four generated GP volatility models are selected for time series classification (MSR, MSS, MSAS and MSAR). Similarly, four generated GP volatility models are selected for moneyness-time to maturity classification (MCR, MCS, MCAS and MCAR) and four generated GP volatility models are selected for global classification, both time series and moneyness-time to maturity classes (MGR, MGS, MGAS and MGAR). Table 5 reports the best generated GP volatility models, using dynamic training-subset selection, relative to *TS* samples, *MTM* classes and both *TS* and *MTM* data.

| TS Models | MSE Total | MTM Models | MSE Total | Global Models | MSE Total |
|---|---|---|---|---|---|
| MSR | 0.002367 (0.003934) | MCR | 0.002427 (0.0037777 | MGR | 0.002034 (0.003501) |
| **MSS** | **0.002076 (0.004044)** | MCS | 0.007315 (0.025811) | MGS | 0.002492 (0.003013) |
| MSAS | 0.002594 (0.003796) | MCAS | 0.002831 (0.004662) | MGAS | 0.001999 (0.003587) |
| MSAR | 0.002232 (0.003782) | **MCAR** | **0.001424 (0.003527)** | **MGAR** | **0.001599 (0.003590)** |

*Table 5: Performance of the generated GP volatility models, using dynamic training-subset selection method, according to MSE total for the TS samples, the MTM classes and both TS and MTM samples.*

Based on the MSE total as performance criterion, the generated GP volatility models MSS, MCAR and MGAR are selected. They seem to be more accurate in forecasting implied volatility than the other models because they have the smallest MSE in enlarged sample.

Table 5 shows that the time series model MSS presents the highest MSE relative to the other models. It seems to be less performing than the moneyness- time to maturity model MCAR and the time series and moneyness- time to maturity model MGAR.

The best generated GP volatility models selected, relative to dynamic training-subset selection method, are compared to the best generated GP volatility model, relative to static training-subset selection method. Results are reported in Table 6.

| Models | MSE total |
|---|---|
| M4S4 | 0,001444 (0,002727) |
| MCAR | 0.001424 (0.003527) |
| **MGAR** | **0.001599 (0.003590)** |

*Table 6: Comparison between best models generated by static and dynamic selection methods for call options*

Comparison between models reveals that the best models generated respectively by static (M4S4) and dynamic selection methods (MCAR and MGAR) present total MSE small and very close. While the generated GP volatility models M4S4 and MCAR have total MSE smaller than the MGAR model, the latest seems to be more accurate in forecasting implied volatility than the other models. This can be explained by the fact that, on one hand, the difference between forecasting errors is small, and on the other hand, the MGAR model is more general than MCAR and M4S4 models because it is adaptive to all time series and moneyness- time to maturity classes simultaneously. In fact, the MGAR model, generated using adaptive-random training-subset selection (ARSS) method, is trained on all time series and moneyness- time to maturity classes simultaneously. Whereas, the MCAR model, generated using adaptive-random training-subset selection (ARSS) method, is trained only on moneyness- time to maturity classes simultaneously; and the M4S4 model, generated using static training-subset selection method, is trained separately on each subset of time series.

As the adaptive- random training subset selection method is considered the best one to generate implied volatility model for call options, it is applied to put options. The decoding of volatility forecasting formulas generated for call and put options as well as their forecasting errors are reported in Table 7.

| Models | Formula | MSE-Total |
|---|---|---|
| GP- call | $\sigma_{GP} = \sqrt{\dfrac{\dfrac{C}{K}}{\left(\dfrac{S}{K}\right)^6 + \left(\dfrac{S}{K}\right)^5 * \tau}}$ | 0.001599 (0.003590) |
| GP-put | $\sigma_{GP} = \Phi\left(\sin\left(\cos\left(\sin\left(-\cos(\sin(\tau)) - 2*\ln\left(\dfrac{P}{K}\right)\right)\right) - \exp\left(\dfrac{S}{K}\right)\right)\right)$ | 0.001539 (0.002158) |

*Table 7: Performance of the best generated GP volatility models for call and put options and their decoding formulas*

A detailed examination of the trees shows that the implied volatilities generated by GP are function of all the inputs used, namely the option price divided by strike price ($\dfrac{C}{K}$ for calls and $\dfrac{P}{K}$ for puts), the index price divided by strike price $\dfrac{S}{K}$ and time to maturity $\tau$. Furthermore, the implied volatilities generated by the MGAR-call and MGAR-put models can't be negative since they are computed using the square root and the normal cumulative distribution functions as the root nodes.
Furthermore, the performance of models is uniform as they present near MSE on the enlarged sample.

### 4.2. Dynamic hedging results:

The performance of the best GP forecasting models is compared to the Black-Sholes model in delta, gamma and vega hedging strategies.
Table 8 reports the average hedging errors for call options using BS and GP models, at the 1-day and 7-days rebalancing frequencies.

| | | | Rebalancing Frequency | | | | | |
|---|---|---|---|---|---|---|---|---|
| | | | 1-day | | | 7- days | | |
| S/K | Hedging strategy | Model | <60 | 60-180 | >=180 | <60 | 60-180 | >=180 |
| <0.98 | Delta hedging | BS | 0,013119 | 0,001279 | 0,000678 | 0,057546 | 0,010187 | 0,005607 |
| | | GP | **0,009669** | **0,001081** | **0,000662** | **0,053777** | **0,009585** | **0,005594** |
| | Gamma hedging | BS | 0,000596 | 0,000732 | 0,000061 | 0,003026 | 0,007357 | 0,000429 |
| | | GP | 0,000892 | 0,002040 | 0,000075 | 0,003855 | **0,001359** | **0,000153** |
| | Vega hedging | BS | 0,000575 | 0,000050 | 0,000039 | 0,000525 | 0,000226 | 0,000099 |
| | | GP | **0,000473** | 0,002035 | 0,004518 | 0,000617 | 0,004642 | 0,040071 |
| 0.98-1.03 | Delta hedging | BS | 0,002508 | 0,000717 | 0,000730 | 0,019623 | 0,005416 | 0,002283 |
| | | GP | **0,002506** | **0,0007** | 0,001725 | 0,020 | **0,0054** | **0,0022** |
| | Gamma hedging | BS | 0,000069 | 0,000018 | 0,000006 | 0,000329 | 0,000169 | 0,000027 |
| | | GP | 0,000377 | 0,000040 | 0,000029 | 0,000727 | **0,000155** | 0,000059 |
| | Vega hedging | BS | 0,000066 | 0,000373 | 0,003294 | 0,000527 | 0,023500 | 0,031375 |
| | | GP | 0,000281 | **0,000013** | **0,000207** | 0,001102 | **0,000147** | **0,000134** |
| >=1.03 | Delta hedging | BS | 0,000185 | 0,000906 | 0,001004 | 0,001602 | 0,006340 | 0,006401 |
| | | GP | **0,000184** | **0,000905** | **0,001** | **0,000840** | **0,005789** | **0,0064** |
| | Gamma hedging | BS | 0,000323 | 0,000047 | 0,000028 | 0,001546 | 0,000386 | 0,000157 |
| | | GP | **0,000028** | 0,000057 | 0,000036 | **0,000227** | 0,000429 | 0,000175 |
| | Vega hedging | BS | 0,000362 | 0,000060 | 0,000052 | 0,001757 | 0,002015 | 0,000247 |
| | | GP | **0,000067** | **0,000057** | **0,00005** | **0,000831** | **0,000864** | **0,000186** |

*Table 8: Average hedge errors of dynamic hedging strategies relative to BS and GP models for call options.*

Results show that the delta hedging performance improves for out-of-the money call options at longer maturities, for at-the-money call options at medium maturities and for in-the money call options at shorter maturities, regardless of model used at daily hedge revision frequency. The best delta hedging performance is achieved using in-the-money short term call options for all moneyness and time to maturity classes, regardless of option model used.

The delta-gamma hedging performance improves for all moneyness classes of call options at longer maturities, regardless of model used at daily hedge frequency (except in-the-money call options using the GP model). The best delta-gamma hedging performance is achieved, for BS model, using at-the-money long term call options for all moneyness and time to maturity classes. However, the best delta-gamma hedging performance is achieved, for GP model, using in-the-money short term call options for all moneyness and time to maturity classes.

The delta-vega hedging performance improves for out-of-the money and in-the-money call options at longer maturities and for at-the-money call options for shorter maturities, for BS model at daily hedge revision frequency. However, the delta-vega hedging performance improves for out-of-the money call options at shorter maturities, for at-the-money call options at medium maturities and for in-the money call options at longer maturities, for GP model at daily hedge revision frequency. The best delta-vega hedging performance is achieved, for BS model, using out-of-the-money long term call options for all moneyness and time to maturity classes. However, the best delta-gamma hedging performance is achieved, for GP model, using at-the-money medium term call options for all moneyness and time to maturity classes.

The percentage of cases where the hedging error of the GP model is less than the BS hedging error is around 59%. In particular, the performance of GP model is better than the BS model on in-the-money call options class. Further, the total of hedging errors relative to GP model is about 21 percent slightly lower than 19 percent relative to BS model.

Table 9 displays the average hedge errors for put options using BS and GP models, at the 1-day and 7-days rebalancing frequencies.

| S/K | Hedging strategy | Model | Rebalancing Frequency | | | | | |
| --- | --- | --- | --- | --- | --- | --- | --- | --- |
| | | | 1-day | | | 7-days | | |
| | | | <60 | 60-180 | >=180 | <60 | 60-180 | >=180 |
| <0.98 | Delta hedging | BS | 0,007259 | 0,002212 | 0,001189 | 0,015453 | 0,013715 | 0,007740 |
| | | GP | 0,064397 | 0,002270 | 0,001256 | 0,016872 | 0,013933 | 0,007815 |
| | Gamma hedging | BS | 0,000107 | 0,000043 | 0,000705 | 0,000383 | 0,000253 | 0,013169 |
| | | GP | 0,000177 | 0,000351 | **0,000676** | 0,000990 | 0,000324 | **0,009201** |
| | Vega hedging | BS | 0,000051 | 0,000715 | 0,000612 | 0,000174 | 0,002995 | 0,008527 |
| | | GP | 0,002800 | **0,000345** | 0,000625 | 0,018351 | **0,000184** | 0,008979 |
| 0.98-1.03 | Delta hedging | BS | 0,007331 | 0,002267 | 0,001196 | 0,170619 | 0,009875 | 0,004265 |
| | | GP | **0,0073** | **0,002219** | **0,001185** | **0,170316** | **0,009715** | **0,004260** |
| | Gamma hedging | BS | 0,003750 | 0,000049 | 0,000027 | 0,032725 | 0,000119 | 0,000119 |
| | | GP | **0,003491** | **0,000031** | **0,000024** | **0,029792** | **0,000113** | **0,000103** |
| | Vega hedging | BS | 0,035183 | 0,000052 | 0,000044 | 0,037082 | 0,000329 | 0,000043 |
| | | GP | **0,004343** | **0,000038** | **0,000043** | **0,037045** | **0,000190** | **0,000041** |
| >=1.03 | Delta hedging | BS | 0,007680 | 0,004469 | 0,000555 | 0,037186 | 0,017322 | 0,011739 |
| | | GP | **0,006641** | **0,004404** | **0,0005** | **0,037184** | **0,017076** | **0,011733** |
| | Gamma hedging | BS | 0,000262 | 0,000204 | 0,000079 | 0,001196 | 0,001319 | 0,000369 |
| | | GP | 0,000548 | 0,000287 | 0,000166 | 0,002034 | 0,001323 | 0,001059 |
| | Vega hedging | BS | 0,000232 | 0,000108 | 0,000025 | 0,000488 | 0,000644 | 0,000270 |
| | | GP | 0,000312 | **0,000080** | **0,00002** | 0,001047 | 0,001186 | **0,000244** |

*Table 9: Average hedge errors of dynamic hedging strategies relative to BS and GP models for put options.*

Results show that the delta-gamma hedging performance improves for all moneyness classes of put options (except in-the-money put options) at longer maturities, for BS model at daily hedge frequency. However, the delta-gamma hedging performance improves for in-the money put options and at-the-money put options at medium maturities and for out-of-the money put options at longer maturities, for GP model at daily hedge revision frequency. The best delta-gamma hedging performance is achieved, for BS model, using at-the-money long term put options for all moneyness and time to maturity classes. However, the best delta-gamma hedging performance is achieved, for GP model, using out-of-the-money long term put options for all moneyness and time to maturity classes.

The delta-vega hedging performance improves for BS using at-the-money and out-of-the-money put options at longer maturities and in-the-money put options for shorter maturities, for BS model at daily hedge revision frequency. However, the delta-vega hedging performance improves for all moneyness classes of put options (except in-the-money put options) at longer maturities, for GP model at daily hedge frequency. The best delta-vega hedging performance is achieved, for BS model, using out-of-the-money long term put options for all moneyness and time to maturity classes. However, the best delta-vega hedging performance is achieved, for GP model, using at-the-money long term put options for all moneyness and time to maturity classes.

The percentage of cases where the hedging error of the GP model is less than the BS hedging error is around 57%. In particular, the performance of GP model is better than the BS model on at-the-money put options class. But, the total of hedging errors relative to GP model is about 50 percent slightly higher than 46 percent relative to BS model.

In summary, the GP model is more accurate in all hedging strategies than the BS model, for in-the-money call options and at-the-money put options. The performance of GP is pronounced essentially in terms of delta hedging for call and put options. The percentage of cases where the delta hedging error of the GP model is less than the BS delta hedging error is 100% for out-of-the money and in-the-money call options as well as for at-the-money and out-of-the-money put options. The percentage of cases where the delta-vega hedging error of the GP model is less than the BS delta-vega hedging error is 100% for in-the-money call options as well as for at-the-money put options. The percentage of cases where the delta-gamma hedging error of the GP model is less than the BS delta-gamma hedging error is 100% for at-the-money put options.

Furthermore, results exhibit that as the rebalancing frequency changes from 1-day to 7-days revision, as the hedging errors increase and vice versa. The option value is a nonlinear function of the underlying, therefore, hedging is instantaneous and hedging with discrete rebalancing gives rise to error. Frequent rebalancing can be impractical due to transactions costs. In the literature, consequences of discrete time hedging have been considered usually in conjunction with the existence of transaction costs, that's why hedgers would like to trade at least frequently as possible. Pioneered by Leland (1985), asymptotic approaches-as are well utilized (Kabanov and Safarian (1997), Ahn et al. (1998), Grandits and Schachinger (2001)). For most moneyness-time to maturity classes, delta-gamma and delta-vega hedging strategies are shown to perform better in dynamic hedging when compared with delta hedging strategy, regardless of model used. The delta-gamma strategy enables the performance of a discrete rebalanced hedging to be improved. The delta-vega strategy corrects partly for the risk of a randomly changing volatility.

## 5. Conclusion

This paper is concerned with improving the dynamic hedging accuracy using generated genetic programming implied volatilities. Firstly, genetic programming is used to predict implied volatility from index option prices. Dynamic training- subset selection methods are applied to improve the robustness of GP to generate general forecasting implied volatility models relative to static training-subset selection method. Secondly, the implied volatilities derived are used in dynamic hedging strategies and the performance of genetic programming is compared to that of Black-Scholes in terms of delta, gamma and vega hedging.

Results show that the dynamic training of GP yields better results than those obtained from static training with fixed samples, especially when applied on time series and moneyness-time to maturity samples simultaneously. Based on the MSE total as performance criterion, three generated GP volatility models are selected M4S4, MCAR and MGAR. However, the MGAR seems to be more accurate in forecasting implied volatility than MCAR and M4S4 models because it is more general and adaptive to all time series and moneyness- time to maturity classes simultaneously.

The main conclusion concerns the importance of implied volatility forecasting in conducting hedging strategies. Genetic programming forecasting volatility makes hedge performances higher than those obtained in the Black-Scholes world. The best GP hedging performance is achieved for in-the-money call options and at-the-money put options in all hedging strategies. The percentage of cases where the hedging error of the GP model is less than the BS hedging error is around 59% for calls and 57% for puts. The performance of GP is pronounced essentially in terms of delta hedging for call and put options. The percentage of cases where the delta hedging error of the GP model is less than the BS delta hedging error is 100% for out-of-the money and in-the-money call options as well as for at-the-money and out-of-the-money put options. The percentage of cases where the delta-vega hedging error of the GP model is less than the BS delta-vega hedging error is 100% for in-the-money call options as well as for at-the-money put options. The percentage of cases where the delta-gamma hedging error of the GP model is less than the BS delta-gamma hedging error is 100% for at-the-money put options.

Finally, improving the accuracy of implied volatility forecasting using genetic programming can lead to well hedged options portfolios relative to the conventional parametric models.